\documentclass[preprint]{elsarticle}

\usepackage{bm}% bold math
\usepackage{lineno,hyperref}
\modulolinenumbers[2]

\usepackage{graphicx}% Include figure files
\journal{Computer Physics Communications}

\usepackage[dvipsnames]{xcolor}
\usepackage{longtable}
\usepackage{graphicx}% Include figure files
\usepackage{dcolumn}% Align table columns on decimal point
\usepackage{bm}% bold math
\usepackage{hyperref}% add hypertext capabilities
\usepackage{float}
\usepackage{color, colortbl}
\usepackage{changes}
\usepackage{amsmath,amssymb}
\usepackage{float}
\usepackage{xcolor}

\usepackage{changes}

\newcommand{\INT}{\int d^3{\bf r}}
\newcommand{\IINT}{\iint d^3{\bf r}\,d^3{\bf r'}}
\newcommand{\vecr}{ {\bf r } }

\definecolor{Gray}{gray}{0.9}

\begin{document}

\title{
Direct Energy Minimization Based on Exponential Transformation in Density Functional Calculations of Finite and Extended Systems
%using  in the Localised Basis Set Approach
}

% -----------------------------------
\author[UI,SPBSU]{Aleksei V.~Ivanov}
\ead{alxvov@gmail.com}
\author[UI]{Elvar \"{O}. J\'{o}nsson}
\author[DTU]{Tejs~Vegge}
\author[UI]{Hannes J\'{o}nsson}
\ead{hj@hi.is}

\address[UI]{Science Institute and Faculty of Physical Sciences, University of Iceland VR-III,107 Reykjav\'{\i}k, Iceland}
\address[SPBSU]{St.\,Petersburg State University, 199034, St.\,Petersburg, Russia}

\address[DTU]{ Department of Energy Conversion and Storage, Technical University of Denmark, DK-2800 Kgs. Lyngby, Denmark}


\newcommand\derA[1]{\frac{\partial #1}{\partial A_{ij} } }

\def\dd#1#2{\frac{\partial#1}{\partial#2}}

\def\bk{{\bf k}}
\def\br{{\bf r}}
\def\bRl{{\bf R}_l}
\def\bR{{\bf R}}
\def\bdmu{{\bf d}_\mu}
\def\bdA{{\bf d}_A}

% -------------------------------

\begin{abstract}
The energy minimization involved in density functional calculations of electronic systems can be carried out using an exponential transformation that preserves the orthonormality of the orbitals. The energy of the system is then represented as a function of the elements of a skew-Hermitian matrix that can be optimized directly using unconstrained minimization methods. An implementation based on the limited memory Broyden-Fletcher-Goldfarb-Shanno approach with inexact line search and a preconditioner is presented and the performance compared with that of the commonly used self-consistent field approach.  Results are presented for the G2 set of 148  molecules, liquid water configurations with up to 576 molecules and some insulating crystals. A general preconditioner is presented that is applicable to systems with fractional orbital occupation as is, for example, needed in the k-point sampling for periodic systems. This exponential transformation direct minimization approach is found to outperform the standard implementation of the self-consistent field approach in that all the calculations converge with the same set of parameter values and it requires less computational effort on average. The formulation of the exponential transformation and the gradients of the energy presented here are quite general and can be applied to energy functionals that are not unitary invariant such as self-interaction corrected functionals. 
\end{abstract}

% \pacs{Valid PACS appear here}% PACS, the Physics and Astronomy
\maketitle

%\tableofcontents

\section{\label{sec: introduction} Introduction}

There are several different approaches for finding optimal orbitals corresponding to the minimum of an energy functional in the context of Kohn--Sham density functional theory (KS-DFT)~\citep{hohenberg1964,kohn1965}. The most commonly used method is based on a self-consistent field (SCF) algorithm consisting of two steps. In the first step and for a given density, one finds eigenvalues and eigenfunctions using an iterative algorithm such as the Davidson algorithm~\citep{Davidson1975} or even direct diagonalization of the full Hamiltonian matrix when the size of the basis set is not too large. 
In the second step, the electron density or Hamiltonian matrix is updated using, 
for example, direct inversion in the iterative subspace (DIIS) 
method~\citep{Pulay1980,Pulay1982}. 
The SCF approach is widely used and has proven to be efficient for both finite (molecules/clusters)
and extended systems, but can, nevertheless, suffer from convergence problems. 
Various density and Hamiltonian mixing schemes have been introduced to address such cases~\citep{Kresse1996,Garza2012}. 
As a result, the user of typical software developed for KS-DFT calculations is often presented with the task of choosing values of various parameters and select between various types of eigensolvers.
Systems with similar chemical and physical properties may even call for different choices.
A further problem of the SCF method in calculations of ground electronic states
is that it may converge on a saddle point of the energy surface rather than a minimum~\citep{Vaucher2017}.    

Another approach to this optimization problem
is based on direct minimization of the energy with respect to the electronic degrees of freedom~\citep{Head-Gordon1988,Gillan1989,Payne1992,Hutter1994,Marzari1997,Ismail-Beigi2000,VanVoorhis2002,VandeVondele2003,Weber2008,Freysoldt2009}.
The challenge then is to incorporate the constraint of orthonormality of the orbitals (the single electron wave functions). 
One way to approach this is to follow the energy gradient projected on the subspace tangent to the orbitals~\cite{Gillan1989,Payne1992}. After such an adjustment of the orbitals within this tangent space, the orthonormality constraints will be violated and, therefore, an explicit orthonormalization of the orbitals needs to be applied  
after each iteration. This approach is often used in calculations with a plane wave basis set. 
Alternatively, when the basis set is compact, as in calculations using linear combination of atomic orbitals,
a unitary transformation can be applied to a set of orthonormal reference orbitals that 
includes all occupied and virtual orbitals, 
and the energy is then minimized by optimizing the elements of the transformation matrix.
The orthonormality constraints will then be satisfied, 
but, due to the constraints imposed by the unitary matrix,
the energy is defined on a curved space.
As a result, minimization algorithms need to be modified to take the curvature into account.
This can be achieved by performing a line search along geodesics~\citep{Edelman1998}. 
Alternatively, the unitary matrix can be parameterized using an exponential transformation
~\citep{Head-Gordon1988,Hutter1994,Ismail-Beigi2000}
in which case the energy becomes a function of the elements of a skew-Hermitian matrix in linear space. 
Well-established, unconstrained minimization strategies can then be applied
including inexact line searches that can give robust convergence.
We will refer to this approach as exponential transformation direct minimization (ETDM). Furthermore, 
it has been used in calculations of molecules using
KS-DFT~\citep{VanVoorhis2002} and previously in the context of 
Hartree-Fock theory~\citep{Head-Gordon1988,Rico1983a, Rico1983b, Douady1980}. 
There, the occupation numbers for the orbitals have been restricted to integers
so that unitary invariance  
with respect to rotation within the space of occupied orbitals is ensured.
Preconditioners to accelerate convergence have been
presented for such systems and found to be important in order to achieve good 
performance~\citep{Head-Gordon1988,VanVoorhis2002}.

In this article, 
a generalization and efficient implementation of the ETDM approach is presented  as well as applications to 
both finite and extended systems.
The method can be applied to systems with fractional occupation, 
for example, where k-point sampling of the Brillouine zone (BZ) is carried out.
The formulation presented here is also applicable  
to energy functionals that are not unitary invariant,
such as self-interaction corrected functionals~\citep{Perdew1981}. 
Tests of the performance of this ETDM implementation and comparison with the SCF method including density mixing are carried out for
the G2 set (a total of 148 molecules), liquid configurations consisting of up to 576 water molecules and several insulating crystals.

The article is organised as follows. 
In section~\ref{sec:theory}, the ETDM method is formulated in a general way 
and equations provided for the derivative of the energy with
respect to the matrix elements in the exponential transformation. 
In section~\ref{sec:algorithms}, an efficient preconditioner is presented, applicable to systems with non-integer occupation numbers,
as well as methods for evaluating the gradient of the energy and ways to choose the search direction as well as step-length in an inexact line-search procedure.
In section~\ref{sec:results}, performance tests are presented with 
comparison to conventional SCF calculations. 
Finally, discussion and conclusions are presented in section~\ref{sec:discussion}.

% ----------------------------------------

\section{\label{sec:theory} General formulation}
In KS-DFT, the energy functional is
\begin{align}
\nonumber
E =
\sum_{i, {\bf k}} f_i(\bk)\INT \frac{|\nabla \phi_{i \bk} ( \vecr )|^2}{2} 
+\INT \rho(\vecr)v_{ext}(\vecr)+\\
+ \frac{1}{2} \IINT  \frac{\rho(\vecr)\rho(\vecr')}{|\vecr - \vecr'| }
+E_{xc}[\rho({\bf r})].
\end{align}
where the $\phi$ are orbitals of the non-interacting electron system that has 
total electron density
\begin{equation}
\rho(\vecr) = \sum_{i, {\bf k}}  f_{i}(\bk) |  \phi_{i \bk} ( \vecr )|^2,
\end{equation}
equal to that of the interacting electron system, 
the $f_{i}({\bf k})$ are orbital occupation numbers for the $k$-th point of the BZ with $0 \le f_{i}({\bf k}) \le 1$,
$v_{ext}(\vecr)$ is the external potential corresponding to electron-nuclei 
interaction, 
and $E_{xc}$ is the exchange-correlation energy.
The orbitals are expanded in terms of a possibly non-orthogonal basis set consisting of $M$ basis functions
\begin{equation}
\phi_{i \bk} (\br) = \sum_{\mu=1}^M O_{\mu i}(\bk) \chi_{\mu \bk}(\br),
\end{equation}
and the task is to find optimal values of the coefficients $O_{\mu i}(\bk)$ 
that minimize the energy $E[\{O(\bk)\}_{\bk}]$ 
subject to the orthonormality constraints:
\begin{equation}
O^{\dagger}({\bf k})S({\bf k})O({\bf k}) = I   \quad \bk \in BZ,
\end{equation}
with $S_{\mu\nu}({\bf k}) = \int \chi^{*}_{\mu \bk}(\br)  \chi_{\nu \bk}(\br)  \,d\br$ being the overlap matrix.

The basis functions for periodic systems are Bloch states and in a localised basis set approach they can be written as
\begin{equation}
\chi_{\mu \bk}(\br) = \frac{1}{\sqrt{N}} \sum_{\bR} \exp(i \bk \cdot \bR) \eta_{\mu} (\br - \bR - {\bf d}_\mu)
\end{equation}
where $\eta_{\mu} (\br - \bR - {\bf d}_\mu)$ is an atomic orbital centered on an atom in the simulated cell. 
The subscript $\mu$ enumerates the atomic orbitals and $\bR$ belongs to the Bravais lattice.
An initial guess for the orbitals is expressed as a linear combination of the basis functions
\begin{equation}
\psi_{m \bk} (\br) = \sum_{\mu=1..M} C_{\mu m}(\bk) \chi_{\mu \bk}(\br) .
\end{equation}

Given an initial guess for the orbitals, $C_{\mu m}(\bk)$, which we will refer to as the reference orbitals, the optimal orbital coefficients $O_{\mu m}(\bk)$ that provide minimal energy can be found through a unitary transformation as
\begin{equation}\label{eq: unit_transf}
O(\bk)=C(\bk)e^{A(\bk)}
\end{equation}
where $A(\bk)$ is a skew-Hermitian matrix, $A(\bk)^{\dagger} = -A(\bk)$.
For a set of $N_k$ vectors used to represent the BZ, 
% \begin{equation}
% A =
% \begin{pmatrix}
% a_{11} & a_{12} & a_{13}& \ldots & a_{1M} \\
% -\overline{a_{12}} & a_{22}& a_{23}&\ldots &a_{2M} \\
% -\overline{a_{13}} & -\overline{a_{23}}&a_{33}&\ldots &a_{2M}\\
% \vdots & \vdots & \vdots & \ddots & \vdots\\
% -\overline{a_{1M}} & -\overline{a_{2M}} & -\overline{a_{3M}} & \ldots & a_{MM}\\
% \end{pmatrix}
% \end{equation}
a set of matrices $\{A(\bk)\}_{\bk}$ is needed. 
For a given set of reference orbitals, a set of 
unitary matrices, $U(\bk) = \exp(A(\bk))$, exists so that the reference orbitals are transformed to the optimal orbitals.
Thus, the ground-state energy of the system is a function of the upper triangular elements of a set of matrices $A(\bk)$, 
%$A$, $a_{ij} = (A)_{ij}$:
%
\begin{equation}
E[n] = E[\{a_{11},\ldots,a_{1M},a_{22}, \ldots,a_{2M}, \ldots, a_{MM} \}_{\bk}] 
\end{equation}
where $a_{ij} = (A)_{ij}$ and ${\bk}$ denotes the set of $N_k$ vectors. 
The real part of the diagonal elements of the matrices are zeros and therefore, the energy is a function of $N_kM^2$ variables.
There are $M(M-1)/2$ real elements and $M(M+1)/2$ imaginary elements for every k-point. The energy needs to be minimized with respect to the real and imaginary parts of the matrix elements $\{a_{ij}(\bk)\}_{i\leq j}$. 
Introducing the derivative 
\begin{equation}
\frac{\partial }{\partial a_{ij}(\bk)} = \frac{1}{2} \left( \frac{\partial }{\partial {\rm Re}(a_{ij}(\bk))} - i\frac{\partial }{\partial {\rm Im}(a_{ij}(\bk))} \right)
\end{equation}
the gradient of the energy can be evaluated as
\begin{equation}\label{eq: der_of_en_1}
\frac{\partial E}{\partial a_{ij}(\bk)}= 
\sum_{\mu\nu} H_{\mu\nu}(\bk)
\frac{\partial\rho_{\mu\nu}(\bk)}{\partial a_{ij} (\bk)}
\end{equation}
where the Hamiltonian matrix is
\begin{equation}
H_{\mu\nu}(\bk) = 
\int d\br\,
\chi^{*}_{\mu \bk}(\br) 
\left(
-\frac{1}{2} \nabla^2 + v(\br)
\right)
\chi_{\nu \bk}(\br) .
\end{equation} 
Here, $v(\br)$ is the single electron Kohn-Sham potential, 
and the density matrix is given in terms of the optimal coefficient matrix as
\begin{equation}
\rho_{\mu\nu}(\bk) = \sum_m f_m(\bk) O_{\mu m}(\bk)  \overline{O}_{\nu m}(\bk) .
\end{equation}
By defining the commutator
\begin{equation}
L_{mk}(\bk) = 
\left[
F(\bk), H(\bk)
\right]_{mk},
\end{equation}
where $H(\bk)$ is the Hamiltonian matrix represented in terms of the optimal orbitals 
\[
H(\bk)_{mk} = \sum_{\mu\nu} \overline{O}_{\mu m} (\bk) H_{\mu\nu} (\bk)  O_{\nu k}(\bk) ,
\]
and $F(\bk)$ is a diagonal matrix with occupation numbers $f_{m}(\bk)$ as diagonal elements,
the derivatives in Eq.~\eqref{eq: der_of_en_1} can be written as
\begin{equation} \label{eq:gradients}
\frac{\partial E}{\partial a_{ij}(\bk)} = \frac{2 - \delta_{ij}}{2} \left( \int_0^1 e^{t A(\bk)} L(\bk) e^{-t A(\bk)}  \,dt \right)_{ji}.
\end{equation}
%Since we are looking 
For the optimal orbitals,
the gradient ${\partial E}/{\partial a_{ij}(\bk)}$
must be zero so
\begin{equation} \label{eq:matrix_eq}
\int_0^1 e^{t A(\bk)} L(\bk) e^{-t A(\bk)}  \,dt = 0, 
\quad \bk \in BZ .
\end{equation}
These non-linear equations can be used to find the skew-Hermitian matrix that provides the energy minimum. 
For the remainder of this article, the $k$-point index ${\bf k}$ is omitted for simplicity.

Eq.~\eqref{eq:gradients} is general and can be applied to an objective function that depends explicitly 
on the orbitals as well as the total density, but then the 
definition of $L$ needs to be be changed accordingly. 
For example, for the Perdew-Zunger self-interaction correction (PZ-SIC)~\citep{Perdew1981}, the matrix $L$ for a single k-point calculation is
\begin{equation}
L_{mk} = 
\left[
F, H
\right]_{mk} + f_k \overline{V}_{km} - f_m V_{mk},
\end{equation}
where $V_{km}$ is a matrix element of the SIC potential:
\begin{align}\label{eq:SIC pot}
& V_{mk} = \sum_{\mu\nu} \overline{O}_{\mu m} V^{k}_{\mu\nu} O_{\nu k},\\
& V_{\mu\nu}^{k}  = \int \chi^{*}_{\mu}({\bf r}) \left[ \int d^3{\bf r'}  \frac{\rho_k(\br')}{|\br - \br'| }  + v_{xc}(\rho_k(\br)) \right]\chi_{\nu}({\bf r}) d{\bf r} .
\end{align}

Equation~\eqref{eq:matrix_eq} can be expanded in a series as
\begin{equation} \label{eq: grad_expansion}
\int_0^1 e^{t A} L e^{-t A}  \,dt =  L + \frac{1}{2!} \left[ A, L\right] + \frac{1}{3!} \left[ A ,\left[ A, L\right]\right] + \ldots
\end{equation} 
If $ \| L \| \gg \frac{1}{2}\| \left[ A, L\right] \| $, then the first term on the right hand side 
can be used to estimate the gradient. 
This limit 
of `small rotations' 
corresponds to the geometric approach used by Van Voorhis and Head-Gordon~\citep{VanVoorhis2002} 
and has also been
used in the context of orbital-density dependent functionals~\citep{Lehtola2014,Borghi2015,Lehtola2016}.
The higher order terms can also be included 
to increase the accuracy of the gradient estimate, but 
each iteration then requires more computational effort.

The minimization procedure is performed with respect to the real and imaginary parts of matrix elements using the energy gradient given by Eq.~\eqref{eq:gradients}
\begin{equation}
\frac{\partial E}{\partial {\rm Re}(a_{ij})} = 2 {\rm Re} \left( \frac{\partial E}{\partial a_{ij}} \right)
\end{equation}
and
\begin{equation}
\frac{\partial E}{\partial {\rm Im}(a_{ij})} = -2 {\rm Im} \left( \frac{\partial E}{\partial a_{ij}}\right).
\end{equation}
Computational algorithms for the evaluation of the the matrix exponential 
and gradient of the energy are presented in Sec.~\ref{subsec: unim_and_grad}
%Appendix~\ref{sec: calc_grad} .
% The index ${\bk}$ is omitted in the following sections for clarity.  

% -----------------------------------------------

\section{\label{sec:algorithms} Algorithms and Computational Parameters}

In order to find the optimal orbitals, $O$, corresponding to minimal energy, the appropriate exponential transformation of the reference orbitals, $C$, 
\begin{equation}
O=Ce^{A}
\end{equation}
needs to be determined.
The reference orbitals can be chosen to be any set of orthonormal orbitals spanned by the basis set and
they are held fixed during the minimization of the energy for a given number of steps while only the matrix $A$ is varied.
The closer the reference orbitals are to the optimal orbitals,
the faster the iterative procedure will converge.

A line search method has been implemented where the $(k+1)$th iteration step is 
\begin{equation}
\vec {a}\,^{(k+1)} = \vec {a}\,^{(k)} + \alpha\,^{(k)}\, \vec{p}\,^{(k)} .
\end{equation}
Here, $\vec {a}\,^{(k+1)}$ is a vector consisting of the real and imaginary part of the upper triangular  elements of
matrix $A$ at the $k$th step of the minimization algorithm,
\begin{equation}
\begin{split}
\vec{a} = 
(&{\rm Re} (a_{12}),\ldots, {\rm Re} (a_{1M}),\\
& {\rm Re} (a_{23}), \ldots, {\rm Re} (a_{2M}), \ldots, {\rm Re} ( a_{M-1M}),\\
& {\rm Im}(a_{11}), {\rm Im}(a_{12}),\ldots, {\rm Im}(a_{1M}),\\
& {\rm Im}(a_{22}), {\rm Im}(a_{23}),\ldots, {\rm Im}(a_{2M}), \ldots,{\rm Im}(a_{MM}) )^T ,
\end{split}
\end{equation}	
and $\vec{p}\,^{(k)}$ is the search direction while  $\alpha^{(k)}$ is the step length. 

% --------------------

\subsection{\label{subsec: search_dir} Choice of search direction}
The search direction can be chosen according to the steepest descent method, 
various Quasi-Newton methods, or nonlinear conjugate gradient (CG) methods.
The calculation of the search direction involves
algebraic operations associated with the particular method plus the 
evaluation of the energy and gradient for the given energy functional. 
The dimensionality of the minimization problem scales as
$NM$, where $N$ is the number of occupied orbitals and $M$ is the number of basis functions. While 
Quasi-Newton methods such as the Broyden-Fletcher-Goldfarb-Shanno (BFGS) algorithm require fewer iterations than limited-memory BFGS (L-BFGS) or CG, the algebraic operations become a bottleneck even for systems of moderate size 
(the BFGS algorithm scales as $\mathcal{O}(N^{2}M^2)$ 
~\citep{Nocedal2006}).  
However, every iteration of the L-BFGS algorithm, in which the approximate inverse Hessian matrix is updated, can be computed with the cost of $\mathcal{O}(mNM)$ operations, where $m$ is 
the number of previous steps stored in memory.  
In the present implementation, the L-BFGS algorithm 
as described in
Ref.~\citep{Nocedal2006} is used and $m=3$ 
in the benchmark calculations. 

% ------
 
\subsection{\label{subsec: step_length}  Choice of step length}
The step length $\alpha^{(k)}$ is chosen in such a way that it satisfies the strong Wolfe conditions~\citep{Wolfe1969,Wolfe1971,Nocedal2006}
\begin{align}
\label{eq: strw1}
& E(\vec {a}\,^{(k)} + \alpha\,^{(k)} \vec {p}\,^{(k)}) \le E(\vec {a}\,^{(k)}) + c_1 \alpha\,^{(k)} \nabla_{\vec {a}} E(\vec {a}\,^{(k)})\cdot \vec {p}\,^{(k)}
\end{align}
and
\begin{align}
\label{eq: strw2}
& |\nabla E(\vec {a}\,^{(k)} + \alpha\,^{(k)} \vec{p}\,^{(k)})\cdot \vec{p}\,^{(k)} |  \le c_2 | \nabla_{\vec{a}} E(\vec {a}\,^{(k)})\cdot \vec {p}\,^{(k)} |
\end{align}
with $0<c_1<c_2<1$. A trial step of $\alpha\,^{(k)}=1$ is always used first to test the conditions. 
After several iterations, a step length of 1 guarantees that the strong Wolfe conditions are satisfied 
in the L-BFGS algorithm~\citep{Nocedal2006}. This is appealing since it reduces the number of energy and gradient calculations which are computationally most intensive in KS-DFT calculations. If $\alpha\,^{(k)}=1$ is not satisfied by the strong Wolfe conditions, then the inexact line search based on the interpolation of the energy along the search direction is used~\citep{Nocedal2006}.
When the energy of the system is evaluated, the 
KS-DFT potential needs to be obtained and, as a result, there is little additional effort involved in evaluating the gradient. Therefore, the energy along the search direction is always interpolated by a cubic function using information about the energy values and gradient at the boundaries of the search interval $[a, b]$. Alongside the strong Wolfe conditions, approximate Wolfe conditions are also checked~\citep{Hager2006}
at the minimum of the interpolated cubic function
\begin{equation}\label{eq: aprw}
(2\delta - 1)\nabla_{\vec{a}} E(\vec {a}\,^{(k)})^T \vec {p}\,^{(k)} \ge
\nabla_{\vec{a}} E(\vec {a}\,^{(k)} + \alpha\,^{(k)} \vec {p}\,^{(k)}) \cdot \vec {p}\,^{(k)} \ge
 \sigma \nabla_{\vec{a}} E(\vec {a}\,^{(k)}) \cdot \vec {p}\,^{(k)},
\end{equation}
% \end{widetext}
%
% \begin{multline}\label{eq: aprw}
% (2\delta - 1)\nabla_{\vec{a}} E(\vec {a}_k)^T \vec {p}_k \ge \\
% \nabla_{\vec{a}} E(\vec {a}_k + \alpha_k \vec {p}_k)^T \vec {p}_k \ge \\
%  \sigma \nabla_{\vec{a}} E(\vec {a}_k)^T \vec {p}_k,
% \end{multline}
and the condition
\begin{equation}\label{eq: aprw2}
 E(\vec {a}\,^{(k)} + \alpha\,^{(k)} \vec {p}\,^{(k)})  \le E(\vec {a}\,^{(k)}) + \epsilon |E(\vec {a}\,^{(k)})| 
\end{equation}
where $\delta < {\rm min} \{0.5, \sigma\}$, $0 < \sigma < 1$ and $\epsilon$ is a small fixed number.
Thus, the line search algorithm is terminated when either the strong Wolfe conditions of Eqs.~\eqref{eq: strw1}-\eqref{eq: strw2} or the approximate Wolfe conditions of Eq.~\eqref{eq: aprw} along with the condition in Eq~\eqref{eq: aprw2} holds.  
The parameter values are set 
to~\citep{Nocedal2006,Hager2006}
\begin{equation}
c_1 = 10^{-4}, c_2 = 0.9, \delta = 0.1, \sigma =0.9, \epsilon=10^{-6}.
\end{equation}

% # this is not valid anymore as we use now good poisson solver.
%A problem which can arise during the line search is that there is noise in the energy evaluations, for example, due to inaccuracy of the Poisson solver. If the magnitude of the noise is comparable with change in the energy during the line search then it is clear that the choice of optimal step based on the interpolation is not applicable. However, for L-BFGS algorithm the step length which equals 1 can be sufficient enough to converge the system near the minima. In our implementation, when gradients are small, for example around $10^{-6}$ eV, a step length of 1 is always chosen and the stability of L-BFGS algorithm is checked. If approximation matrix to the inverse Hessian is not positive definite then one step along a steepest decent direction is performed after which the L-BFGS algorithm is used again. However, for practical calculations, a convergence threshold of $10^{-5}$ eV for the energy gradients provides good accuracy for the minumum energy and resulting atomic forces.

% -----------

\subsection{\label{subsec: precond} Preconditioning}

A preconditioner speeds up convergence of this iterative algorithm.
It is constructed as the inverse of an approximate Hessian matrix
that can be obtained by taking the derivative of a linear expansion of the gradient~(Eq.~\eqref{eq:gradients}) with respect to the skew-Hermitian matrix, and neglecting first order derivatives of the effective potential. Neglecting the first order derivatives of the effective potential means that all explicit contributions from the Hartree-Exchange-Correlation kernel are neglected.
For the real valued case, the Hessian can be approximated as
\begin{align}
\nonumber
\frac{\partial^{2} E}{\partial a_{ij} \partial a_{lm}} \approx\,
 & \delta_{il}H_{jm} (f_{l} + f_{i} - f_{j} - f_{m}) \\
\nonumber
+&\delta_{jl} H_{im} (f_{m} + f_{i} - f_{l} - f_{j})\\
\nonumber
+& \delta_{jm} H_{li} (f_{m} - f_{i}  - f_{l} + f_{j}) \\
\nonumber
+& \delta_{im} H_{lj} (f_{l} - f_{m} - f_{i} + f_{j}) \\
+& \beta_{ij} \delta_{il} \delta_{jm}
\end{align}
where the matrix $\beta_{ij}$ must be chosen according to the following two principles: (1) the approximate Hessian must be positive definite,
and (2) it must provide a good estimate of the true Hessian along the search direction such that a step size of 1 satisfies the strong Wolfe conditions.

If the orbitals are chosen as eigenvectors of the Hamiltonian then the approximate Hessian is diagonal
\begin{equation}\label{eq: hessian}
\frac{\partial^{2} E}{\partial^{2} a_{ij}} = -2 (\epsilon_{ii} - \epsilon_{jj} ) (f_{i} - f_{j})
 + \beta_{ij} .
\end{equation}
If one keeps contributions from the Hartree-Exchange-Correlation kernel then the Hessian matrix is not diagonal and the inversion of this matrix will require considerable computational effort.

The first term on the right hand side coincides with 
the preconditioner that has previously been used for molecular systems with integer occupation numbers~\citep{Head-Gordon1988}. 
There, an extra term was added in cases of degeneracy, 
$\epsilon_{ii} = \epsilon_{jj}$,
but here the initial approximation of the Hessian in the L-BFGS algorithm~\citep{Nocedal2006} $\beta_{ij}$ is  used.
Since the approximate Hessian is diagonal, the preconditioner is simply
\begin{equation}\label{eq:prec_theor}
P_{ij} = \frac{1}{ -2 (\epsilon_{ii} - \epsilon_{jj} ) (f_{i} - f_{j})
 + \beta_{ij} } .
\end{equation}
In the present implementation, the preconditioner is updated iteratively and for 
iteration k it is
\begin{equation}\label{eq:prec_3}
P^{(k)}_{ij} = \frac{1}{ - 2(1 - \gamma)  (\epsilon_{ii} - \epsilon_{jj} ) (f_{i} - f_{j})
 + \gamma \beta^{(k)} },
\end{equation} 
where  
%a scaling factor 
\begin{equation}
\label{eq:betaij}
\beta^{(k)} = \frac{\| \nabla_{\vec{a}} E(\vec {a}\,^{(k)})  - \nabla_{\vec{a}} E(\vec {a}\,^{(k-1)})  \|^2}{(\vec{a}\,^{(k)} - \vec{a}\,^{(k-1)} )\cdot (\nabla_{\vec{a}} E(\vec {a}\,^{(k)}) - \nabla_{\vec{a}} E(\vec {a}\,^{(k-1)})  )} .
\end{equation}
The parameter $\gamma$ in Eq.\eqref{eq:prec_3} is a number that determines the mixing of the two approximate Hessians: the one obtained from a linear expansion of the gradient, Eq.\eqref{eq: grad_expansion}, and the one based on the LBFGS 
estimate, Eq.\eqref{eq:betaij}. 
In the calculations presented here, $\gamma = 0.25$ was found empirically to give a good compromise between the rate of convergence and robustness. 
When k-point sampling is included for the periodic systems, $\beta^{(k)}$, needs to be multiplied by the numerical weight of the corresponding k-point. 
Eq.~\eqref{eq:prec_3} is used as an initial inverse Hessian at each iteration of the L-BFGS algorithm.

With this preconditioner, a step length of 1 is almost always accepted and it works well for both finite and extended systems. 
It is used for both the real and imaginary parts of the skew-Hermitian matrix. 
We note that the eigenvalues in Eq.~\eqref{eq:prec_3} are not updated at every iteration of the minimization algorithm but only at the beginning,
thereby avoiding
the costly diagonalization of the Hamiltonian matrix at each step.
% -------------------------------------
 
\subsection{\label{subsec: unim_and_grad} Evaluation of the matrix exponential and energy gradient}
%\subsubsection{The Matrix Exponential}
  
The evaluation of the exponential of the skew-Hermitian matrix, $\exp(A)$, 
is carried out using eigendecomposition of $iA$. Let $\Omega$ be a diagonal, real-valued matrix with elements corresponding to the eigenvalues of the matrix $iA$ and let $U$ be a column matrix of the eigenvectors of $iA$. Then the matrix exponential of $A$ is
\begin{equation}\label{eq: ed}
\exp(A) = U \exp(-i\Omega) U^{\dagger}.
\end{equation}
This computation requires diagonalization of a $M\times M$ matrix and
becomes a computational bottleneck for large systems. 
However, for unitary invariant energy functionals (such as Kohn-Sham functionals), Hutter {\it et.al.}~\citep{Hutter1994} have shown that  
$A$ can be parametrised without loss of generality as
\begin{equation}
A = 
\begin{pmatrix}
0 & A_{ov} \\
-A_{ov}^{\dagger} & 0
\end{pmatrix},
\end{equation} 
where $A_{ov}$ is a $N\times (M - N)$ matrix (N - number of occupied states) 
and the matrix exponential can be calculated as
%~\citep{Hutter1994}
% \begin{widetext}
\begin{equation}\label{eq: uinv}
\exp(A) = 
\begin{pmatrix}
{\rm cos}(P) & P^{-1/2} {\rm sin}(P^{1/2}) A_{ov}\\
-A_{ov}^{\dagger} P^{-1/2} {\rm sin}(P^{1/2}) & I_{M-N} + A_{ov}^{\dagger}  {\rm cos}(P^{1/2} - I_N) P^{-1} A_{ov} )
\end{pmatrix},
\end{equation}
% \end{widetext}
where $P = A_{ov} A_{ov}^{\dagger}$. 
In this case the computational effort scales as $\mathcal{O}(N^{2} M)$.

An alternative and more general approach is provided by the scaling and squaring
algorithm based on the equation
\begin{equation}\label{eq: ss}
\exp(A) = \exp(A/2^m)^{2^m}
\end{equation}
and on $[q, q]$ P\'{a}de approximant to the matrix $\exp(A/2^m)$, where $m$ and $q$ are positive integer constants~\citep{Moler2003}. 
The algorithm of Al-Mohy and Higham is used here~\citep{AlMohy2009a,SciPy}.
The two approaches are compared in the benchmark calculations presented below.
% ------

%\subsubsection{Energy Derivatives}
If the matrix exponential is evaluated using the eigendecomposition of $iA$, 
then one can calculate the gradient of the energy using the matrices $U$ and $\Omega$ as
\begin{equation}
G^{T} = U \left( \left( U^{\dagger} LU \right) \otimes D \right)U^{\dagger} ,
\end{equation}
where the matrix $D$ is
\begin{equation}
D_{ij} = \frac{e^{-i(\Omega_{ii} - \Omega_{jj})} - 1}{i(\Omega_{ii} - \Omega_{jj})} 
\end{equation}
and the matrix G is
\begin{equation}
 G_{ij} = \frac{\partial E}{\partial a_{ij}} \frac{2}{ (2 - \delta_{ij})} .
\end{equation}

However, due to the sparsity of the matrix $A$ and if the norm is $\| A\| \ll 1$,  the gradients can be evaluated
more efficiently using only the first term on the 
right hand side of Eq.~\eqref{eq: grad_expansion}
\begin{equation}\label{eq: app_grad}
    G \approx L^T .
\end{equation}
If the norm of the matrix $A$ is larger than 1, then the reference orbitals can be updated $C \leftarrow C\exp(A)$ in which case $A \leftarrow 0$ and then Eq.~\eqref{eq: app_grad} can be used. 
Namely, during the iterative process,
\begin{equation}
    O^{(k)} = C \exp(A^{(k)})
\end{equation}
check if $ \| L^{(k)} \| \ge \epsilon \| \left[ A^{(k)}, L^{(k)}\right] \| $ then set $C' = C \exp(A^{(k)})$, and continue with
\begin{equation}
    O^{(k+1)} = C' \exp(A^{(k+1)}).
\end{equation}
It is found that $\epsilon=3$ provides a reasonable estimate. However, in order to avoid an additional calculation of a commutator between $A$ and $L$, one can update the reference orbitals at a regular interval (for example, at every 20th. iteration).
The change of the reference orbitals should be followed by a transformation of the gradient vectors, as stored in memory for quasi-Newton algorithms, if they are to be used in following steps. However, in the implementation used for the numerical tests in this work, the memory of the L-BFGS algorithm is instead erased after an update of the reference orbitals. This can be beneficial in cases where the orbitals are near stationary points which are not the minimum.

For small systems, the performance is similar for the various methods for evaluating the matrix exponential and energy gradient since the calculation of the effective Kohn-Sham potential and the total energy then dominates the computational effort.
For larger systems, a difference in performance becomes evident, as illustrated
below for configurations of liquid water with up to 576 molecules.
% ------

\subsection{Implementation and parameter values}
We have implemented the ETDM algorithm using a numerical localized atomic basis set 
%representation of the orbitals 
and the projector augmented-wave formalism (PAW)~\citep{Blochl1994}
to take into account the frozen, inner electrons of the atoms within the 
open-source GPAW software~\citep{Enkovaara2010}. 
An SCF algorithm based on the eigendecomposition of the Hamiltonian in a localised atomic basis set representation is already available there and is frequently used in KS-DFT calculations~\citep{Larsen2009}. 
To compare the efficiency of the two approaches, single-point ground-state energy calculations are performed for the G2~\citep{Curtiss1997} data set of
small molecules, five ionic solids, as well as liquid water configurations
including 32, 64, 128, 256, 384 and 576 molecules subject to periodic boundary conditions. 
The double-zeta polarized basis set (which is the default basis set in GPAW) and the generalized gradient approximation (GGA) parametrized by Perdew-Burke-Ernzerhof~\citep{Perdew1996} is used. 
An initial guess for the orbitals is taken to be the eigenvectors of the 
Hamiltonian obtained from a superposition of atomic densities.

Convergence is considered achieved for both the SCF and the ETDM methods when the inequality
\begin{equation}\label{eq: crit}
\frac{1}{N_e} \sum_{i = 1}^{N_b}  \int d\, \br f_i |\hat H_{KS} \psi_i(\br) - \sum_{j=1}^{N_b}\lambda_{ij} \psi_j(\br)|^{2} \ < \ 10^{-10} ~{\rm eV}^{2}  
\end{equation}
is satisfied. In the equation above, the $\lambda_{ij}$ are Lagrange multipliers and for an SCF algorithm this is a diagonal matrix. 
$N_b$ is the number of occupied orbitals.
Default values in GPAW are used, for example the Pulay density mixing parameters.
We note that in cases where the SCF method fails to converge, it could in principle be made to converge by using, for example, other, non-default values of the density mixing parameter. Failure to reach convergence here means that 
convergence is not obtained in the
default maximum number of iteration steps, which is 333.

% -----------------------------------------------------

\section{\label{sec:results}Results}

% -------------

\subsection{Molecules}

\begin{figure*}
\includegraphics[width=1.0\textwidth]{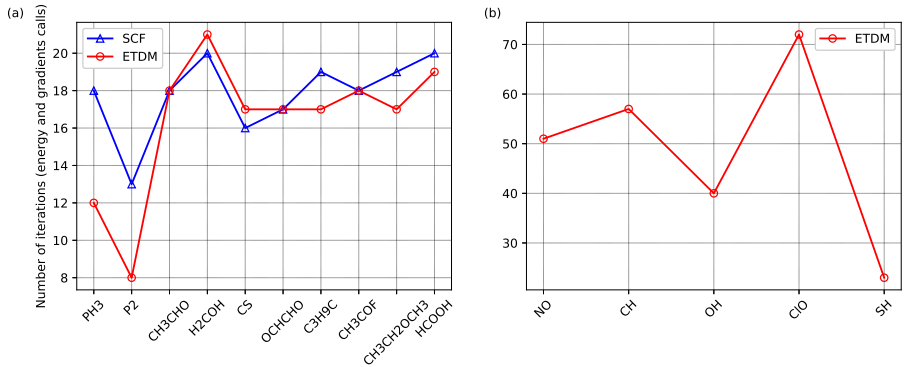}% Here is how to import EPS art
\caption{\label{fig: fig1}
(a) Number of SCF iterations and energy/gradient evaluations in the exponential transform direct minimization needed to reach convergence according to criterion Eq.~\eqref{eq: crit} for a representative set of 10 molecules from the G2 set. (b) Energy/gradient evaluations in the exponential transform direct minimization for the molecules for which the SCF method failed to converge.
}
\end{figure*}

The average number of energy and gradient evaluations for the ETDM method and the average number of energy and diagonalization calculations for the SCF method are presented in Table~\ref{tab: ks_iters_molecules} and Fig.~\ref{fig: fig1}. 
The ETDM method converges for all the 148 molecules in the G2 set using the parameter values
specified in Sec. 3.
The SCF method, however, fails to converge for five of the molecules:  CH, SH, ClO, NO, and OH. 
These five molecules are also challenging for the ETDM method as it requires more iterations to reach convergence there than the average for the whole G2 set (see Fig.~\ref{fig: fig1}).
For the molecules where SCF converges, it requires a similar number of iterations as ETDM. 
On average  18 and 17 iterations are required by the SCF and ETDM methods, respectively.  

The reason for the lack of convergence for SCF and slow convergence of ETDM in the five problematic cases could be the presence of 
nearby saddle points or near-degenerate higher energy states.
In the SCF calculations, the orbitals obtained from the diagonalization of the Hamiltonian matrix at subsequent iterations can `jump' between different energy surfaces or oscillate around a saddle point. Analogous convergence issues for the DIIS method have been reported for the G2 molecular set and transition metal complexes~\citep{VanVoorhis2002}.

% ----  Table -----
\begin{table}[H]
\caption{\label{tab: ks_iters_molecules}
 Comparison of the performance of the 
 exponential transform direct minimization, ETDM, and self-consistent field, SCF, methods
 for the G2 set of molecules (a total of 148 molecules). 
 The average number of energy and gradient evaluations is reported for the former method, but the average number of energy and diagonalization calculations for the latter (in both cases denoted e/g(d)). 
 In the column labeled ETDM$^\ast$, the five molecules for which the SCF calculations did not converge are excluded.}
% \begin{ruledtabular}
\begin{center}
\begin{tabular}{lccc}
\hline
\\
     & SCF & ETDM & ETDM$^\ast$\\
\hline 
\\
%{\bf G2 set} & & \\
average e/g(d) &  18 & 17 & 16 \\
% \hline
min e/g(d) &  12 &  6 &  6\\
% \hline
max e/g(d) &  26  & 72 &  25 \\  
did not converge &5& - &  - \\
  \\
\hline 
%{\bf G2* set} & & \\
%\hline
%average e/g(d) &  18 & 16  \\
% \hline
%min e/g(d) &  12 &  6\\
% \hline
%max e/g(d) &  26 & 25  \\ 
\end{tabular}
\end{center}
% \end{ruledtabular}
\end{table}
% -------------------

For these small molecules, the evaluation of the matrix exponential and energy gradient, i.e. the diagonalization of the Hamiltonian matrix, is not the dominant computational effort. 
The various algorithms presented in Sec~\ref{subsec: unim_and_grad} therefore involve similar computational effort.
%in the calculations of the G2 data set.
%and the computational time is similar to the SCF calculations.

% ------------------ Figure 1 ------------------

% -----------------------------------------------

\subsection{Periodic Systems}
\begin{figure}
\begin{center}
\includegraphics[width=0.7\columnwidth]{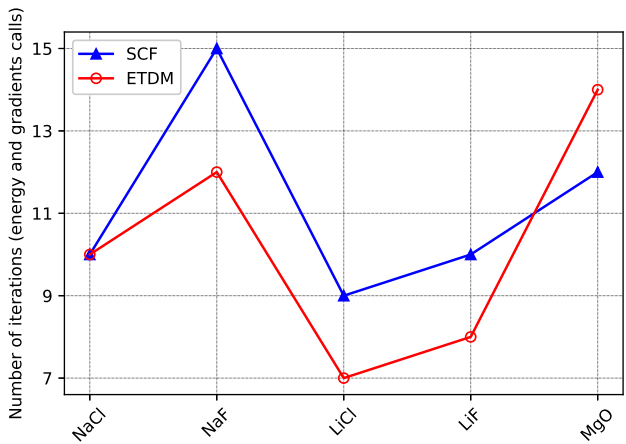}% Here is how to import EPS art
\caption{\label{fig: fig2}
Number of SCF iterations and energy/gradient evaluations in the exponential transform direct minimization needed to reach convergence according to criterion Eq.~\eqref{eq: crit} for NaCl, NaF, LiCl, LiF and MgO crystals.
%Comparison of the number of iterations needed in the SCF calculations and energy-gradient evaluations in the direct minimization calculations needed to reach the convergence as defined by equation~\eqref{eq: crit} for the five ionic crystals. 
}
\end{center}
\end{figure}

As examples of extended systems subject to periodic boundary conditions, 
calculations have been carried out for
five crystalline solids: NaCl, NaF, LiCl, LiF and MgO.
%Ionic solids are considered as a first example of periodic calculations.
A cubic unit cell is chosen consisting of 8 atoms and $\Gamma$-centered $3\times3\times3$ Monkhorst-Pack meshes are used for the BZ sampling. 
The lattice constants are set to the optimal values obtained 
%at the GGA level of theory from Ref.\citep{Staroverov2004}, using the PBE functional. 
from PBE calculations~\citep{Staroverov2004}.
The number of iterations required to reach convergence is presented in Fig.~\ref{fig: fig2}.
The results show that the ETDM and the SCF algorithms have similar rate of convergence for these systems. 
This is an important test of the 
%demonstrates the universality of the 
preconditioner given in equation~\eqref{eq:prec_3} 
and shows that it is suitable for solids as well 
as molecules.

% --------  Figure 2  ---------
% -------------------------------

\begin{figure}
\begin{center}
\includegraphics[width=1.0\columnwidth]{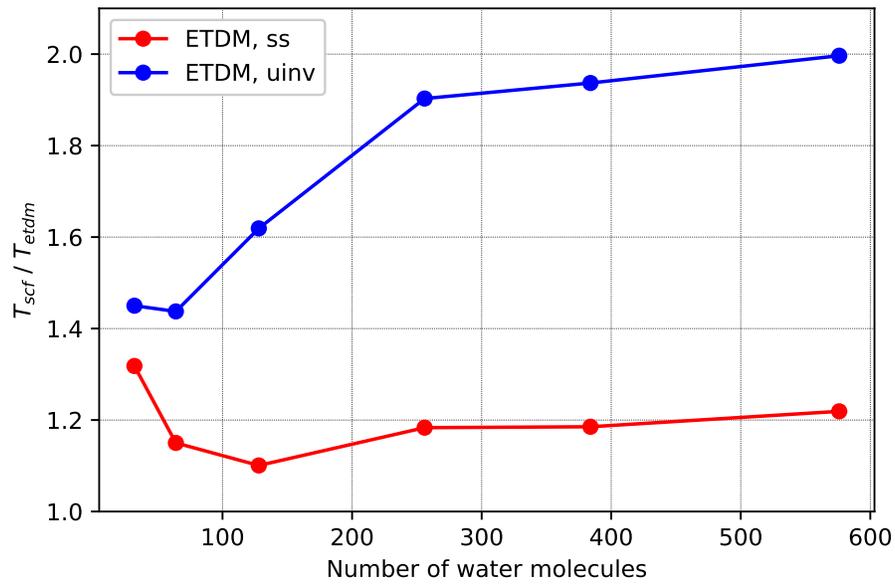}
\caption{\label{fig: fig3}
Ratio of the CPU time used by the 
SCF method and the exponential transform direct minimization, ETDM, method based on 
%one of two different algorithms: 
either the scaling and squaring algorithm, 
Eq.~\eqref{eq: ss} (ss, red curve) or the evaluation of the matrix exponential by diagonalization, Eq.~\eqref{eq: ed}, (uinv, blue curve), as a function of the number of water molecules in liquid configurations subject to periodic boundary conditions. 
For the largest system, the direct minimization based on matrix diagonalization outperforms the SFC method by a factor of two
while the implementation based on the scaling and squaring algorithm is 20\% faster than SCF.
}
\end{center}
\end{figure}

Tests were also carried out for another set of extended systems representing snapshots of liquid water.
The systems contain 32, 64, 128, 256, 384 and 576 water
molecules subject to periodic boundary conditions.
%The system of increased number of water molecules subject to periodic boundary conditions is considered as the second example. 
The efficiency of the two approaches for evaluating the matrix exponential in the ETDM method discussed in Sec~\ref{subsec: unim_and_grad} is compared, also in 
relation to SFC, and reported in Fig.\ref{fig: fig3}. 
One of the approaches is based on Eq.~\eqref{eq: uinv} and makes use of the fact that the energy is invariant with respect to unitary rotations of the occupied orbitals. 
In this case, the computation of the matrix exponential requires diagonalization of an $N \times N$ matrix and involves less computational time as compared to the SCF algorithm where the first $N$ eigenvectors of a $M\times M$ Hamiltonian matrix need to be calculated. 
The other approach, the scaling and squaring algorithm Eq.\eqref{eq: ss}, is more general and does not rely on the parameterization of the skew-Hermitian matrix based on Eq.~\eqref{eq: ss}. For dense matrices, this approach is generally slower than the one based on eigendecomposion of the skew-Hermitian matrix Eq.~\eqref{eq: ed}, but for sparse matrices this algorithm can outperform the eigendecomposion approach. The energy gradient is calculated according to Eq.~\eqref{eq: app_grad}.
%
%------------------  Figure 3 -------------------
%------------------------------------------------

The ratio of the CPU time required by calculations using the SCF method
and the ETDM method is shown as a function of the number of water molecules in Fig.~\ref{fig: fig3}. 
When the matrix exponential is evaluated using Eq.~\eqref{eq: uinv}
the ETDM method outperforms SCF by a factor of two if more than 200 water molecules are included in the system. 
Also, the more general implementation of ETDM using scaling and squaring, Eq.~\eqref{eq: ss}, is faster than SCF by 20\% for these relatively large 
systems. 
It 
has the advantage of being applicable to energy functionals lacking unitary 
invariance, unlike the SCF algorithm.
 
%---------------------------------------------

\section{\label{sec:discussion}Discussion and Conclusion}
The main advantage of the ETDM implementation presented here, based on a general preconditioner, L-BFGS algorithm and inexact line search is robustness. For small molecules
the computational effort is similar to the standard SCF approach when the latter converges, but the ETDM is found to converge for all the molecules in the G2 set with the same set of parameter values, a set that also works for extended liquid configurations and 
insulating solids.
This demonstrates the transferability of the ETDM algorithm as implemented here.
For the large systems considered here, liquid water configurations with 200 and up to 576 molecules, 
the ETDM outperforms the direct SCF method up to by a factor of two
when special parametrization of skew-Hermitian matrix is used 
and by around 20\% when the more general scaling and squaring method is used. 
The latter is more general and can be applied to any type of orbital 
dependent energy functional
 such as self-interaction corrected functionals~\citep{Perdew1981}.

 The ETDM method involves minimization of the energy with respect to the elements of a skew-Hermitian matrix and, therefore, the number of degrees of freedom scales as $M^2$, where $M$ is the number of basis functions.  
 However, for energy functionals that are unitary invariant with respect to the occupied orbitals, the skew-Hermitian matrices can be parametrized using $N\times (M-N)$ degrees of freedom,\citep{Hutter1994} where $N$ is the number of occupied orbitals. 
 Therefore, taking into account the sparsity of the matrices, the algorithm can be implemented in such a way that the computational effort scales as $\mathcal{O}(N^2M)$. 
 The scaling and squaring algorithm for evaluating the matrix exponential 
 is not as efficient but is more generally applicable and can still outperform the SCF method  
 as was found for the large liquid water configurations.

Future work will involve generalization of the ETDM method to finite temperature KS-DFT, i.e. thermal smearing, 
where an additional inner loop for variational optimization of the occupation numbers is included~\citep{Serrano2013},
analogous to the direct minimization method used in ensemble DFT~\citep{Marzari1997}.
This is needed for calculations of metallic systems.  
A more efficient preconditioner could also likely be developed, 
especially for orbital density dependent functionals.
Finally, we point out that the ETDM method is also useful in other types of electronic structure calculations, 
such as studies of excited states~\cite{Levi2020fd,Levi2020}.

% ---------------------- 
 
\section{Acknowledgement}
The authors thank Gianluca Levi for fruitful discussion and valuable comments on the manuscript.
This work was supported by the University of Iceland Research Fund and the Icelandic Research Fund (grant no. 174082-053). AVI is supported by a doctoral fellowship from the University of Iceland. HJ, AVI and E\"OJ thank the Department of Energy Conversion and Storage at the Technical University of Denmark for hospitality during an extended visit and access to computational resources.

\bibliography{main.bib}

\end{document}